\documentclass[aps,prb,reprint,twoside,showpacs]{revtex4-1}
\usepackage{amsfonts}
\usepackage{amsmath}
\usepackage{amssymb}
\usepackage{graphicx}
\usepackage{subfigure}
\usepackage{CJK}
\usepackage{mathptmx}
\usepackage[breaklinks,colorlinks,linkcolor=blue,citecolor=blue,urlcolor=black,bookmarks=false]{hyperref}
\usepackage{placeins}

\begin{document}
\begin{CJK*}{Bg5}{bsmi}
\title{Towards superlattices: Lateral bipolar multibarriers in graphene}%
\author{Martin Drienovsky}%
\email{martin.drienovsky@physik.uni-regensburg.de}
\affiliation{Institut f\"ur Experimentelle und Angewandte Physik, Universit\"at Regensburg, 93040 Regensburg, Germany}%
\author{Franz-Xaver Schrettenbrunner}
\affiliation{Institut f\"ur Experimentelle und Angewandte Physik, Universit\"at Regensburg, 93040 Regensburg, Germany}%
\author{Andreas Sandner}
\affiliation{Institut f\"ur Experimentelle und Angewandte Physik, Universit\"at Regensburg, 93040 Regensburg, Germany}%
\author{Ming-Hao Liu (¼B©ú»¨)}
\affiliation{Institut f\"ur Theoretische Physik, Universit\"at Regensburg, 93040 Regensburg, Germany}%
\author{Fedor Tkatschenko}
\affiliation{Institut f\"ur Theoretische Physik, Universit\"at Regensburg, 93040 Regensburg, Germany}%
\author{Klaus Richter}
\affiliation{Institut f\"ur Theoretische Physik, Universit\"at Regensburg, 93040 Regensburg, Germany}%
\author{Dieter Weiss}
\affiliation{Institut f\"ur Experimentelle und Angewandte Physik, Universit\"at Regensburg, 93040 Regensburg, Germany}%
\author{Jonathan Eroms}%
\affiliation{Institut f\"ur Experimentelle und Angewandte Physik, Universit\"at Regensburg, 93040 Regensburg, Germany}%

\date{\today}%

\begin{abstract}
We report on transport properties of monolayer graphene with a laterally modulated potential profile, employing striped top gate electrodes with spacings of $100$ nm to $200$ nm. Tuning of top and back gate voltages gives rise to local charge carrier density disparities, enabling the investigation of transport properties either in the unipolar (nn$^\prime$) or the bipolar (np$^\prime$) regime. In the latter, pronounced single- and multibarrier Fabry-P\'erot (FP) resonances occur. We present measurements of different devices with different numbers of top gate stripes and spacings. The data is highly consistent with a phase coherent ballistic tight binding calculation and quantum capacitance model, whereas a superlattice effect and modification of band structure can be excluded.
\end{abstract}

\maketitle
\end{CJK*}

\section{Introduction}
As one of the most exciting topics in condensed matter physics \cite{novoselov2012roadmap}, monolayer graphene (MLG) provides a unique combination of striking mechanical, as well as electronic properties, such as the zero gap energy spectrum and linear dispersion of charge carriers (Dirac fermions), leading to the half-integer quantum Hall effect \cite{novoselov2005, novoselov2007room} or relativistic phenomena on a mesoscopic scale (Klein tunneling\cite{klein1929reflexion,dombey1999seventy,katsnelson2006chiral}, electron lensing\cite{cheianov2007focusing}).
In recent years, there has been a vivid discussion on the issue of transport through bipolar junctions giving rise to some thought provoking theoretical approaches\cite{cheianov2006selective, zhang2007nonlinear,shytov2008klein, abanin2007quantized} and experiments\cite{gorbachev2008conductance, ozyilmaz2007electronic, williams2007quantum}. Consequently, different intriguing situations such as Klein tunneling and collimation at a single pn-junction or transmission through a bipolar barrier in the ballistic regime have been covered in scientific publications and revealed resonant behavior of conductance in analogy to optical Fabry-P\'erot (FP) cavities \cite{young2009quantum,stander2009evidence,pereira2010klein}.

Taking a step further, efforts have been taken to investigate the concept of superlattices \cite{esaki1970superlattice} for graphene, employing artificial regular inhomogeneities to the monolayer, experimentally attainable either by gating or doping. For one-dimensional periodic potentials in graphene, theory predicts conductance oscillations corresponding to the emergence of additional Dirac points and the appearance of van Hove singularities in the density of states \cite{park2008anisotropic, abedpour2009conductance, brey2009emerging, barbier2010extra}. Another experimental approach towards superlattices are hexagonal boron nitride (hBN)/graphene heterostructures, where a moire pattern leads to an artificial band structure \cite{yankowitz2012hbnsuperlattice}.

Here we focus on the former approach, addressing finite multibarrier systems \cite{tsu1973tunneling}. Ballistic transmission over a couple of periods is essential for the observation of an artificial band structure. This is why the mean free path and phase coherence length of the charge carriers drastically limit the effect of a superlattice and samples with high charge carrier mobility $\mu$ are needed. Recent works on high-$\mu$ suspended graphene devices \cite{rickhaus2013ballistic, grushina2013ballistic, oksanen2013single}, implementing a single, tunable pn-junction, show stunning resonant behavior. With respect to multiple lateral barriers, suspended graphene is not a viable approach, as the large distance between the graphene-layer and the (buried) gate electrodes causes a significant smoothing of the junction potential.

In order to overcome the problem of blurred potential steps and to provide a locally sharp potential profile at the bipolar junctions, an array of narrow electrode stripes on top of a thin dielectric, as has been introduced for a single bipolar ballistic potential barrier\cite{young2009quantum}, is suitable. We focus our research on locally gated graphene with varying number of bipolar barriers and different spacing in order to characterize the impact of changing parameters on the resistance of the monolayer. A recent publication on a similar lateral multibarrier setup\cite{dubey2013tunable} attributes the occurring resonant behavior of the resistance in the bipolar regime to a superlattice effect in the low-diffusive limit, even though the experimental elastic mean free path does not exceed one lattice period. We reproduce the reported resonant features and provide extended data for varying experimental parameters that can be explained within a consistent Fabry-P\'erot resonance model, without resorting to a superlattice effect.

\section{Electronic transport through multibarriers in graphene}

\subsection{Device fabrication}

\begin{figure}[t]
\includegraphics[width=0.95\columnwidth]{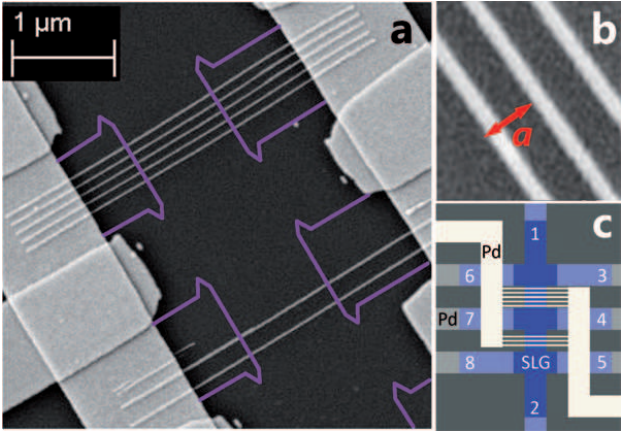}
\caption{\textbf{Device geometry.} SEM image (a) of device B with 5 and 2 top gate stripes. The graphene hallbar is buried under the dielectric, here visualized by the violet contour. Zoom-in (b) of the locally gated area of three Pd stripes of device A (region A3) with a pitch of $a_{\rm A3}=100$ nm. (c) Schematic of the sample geometry of all devices used in this publication employing a MLG hallbar (blue), contacted by a $40$ nm thick Pd fine wiring (light grey). The width of the transport channel is $1$ $\mu$m, as well as the spacing of the labeled voltage probes. The striped Pd top gate electrode of $15$ nm height and the top gate leads (white) are separated from the graphene by a $\sim 15$ nm thin Al$_2$O$_3$ dielectric.}
\label{pic:device}
\end{figure}

Here we present devices where graphene has been exfoliated micromechanically on SiO$_2$ \cite{novoselov2004} (devices A and R), or alternatively, transferred on a hBN\footnote{We used commercially available hBN powder.}/SiO$_2$ substrate (device B) using a PVA/PMMA based method similar to Ref.\ \onlinecite{dean2010hbntransfer}. Hallbars were etched with O$_2$/Ar (50:50) plasma at 30 mTorr and 50 W. Al$_2$O$_3$ was used as a top gate dielectric, built up by a thin film of electron beam evaporated and oxidized aluminum, serving as a thin seed-layer for the following atomic layer deposition (ALD). In total, the top gate oxide does not exceed a height of $15$ nm. The contacts and the patterned top gate electrode were exposed by electron beam lithography and electron beam evaporation of Pd. Experiments were carried out using an ac lock-in four point measurement setup in a $^4$He cryostat ($T = 1.3 \ldots 200$ K) and $B_{\max}=14$ T. Figure \ref{pic:device} depicts the device geometry.

We focus on data of three samples that provide regions with different numbers of top gate stripes and spacing (in the following A3, A5, B2, B5, R8, labels corresponding to number of top gate stripes) and compare them to our ballistic transport simulation.

\subsection{Electronic Transport}

\subsubsection{Preliminary considerations}

\begin{figure}[h]
\includegraphics[width=1.00\columnwidth]{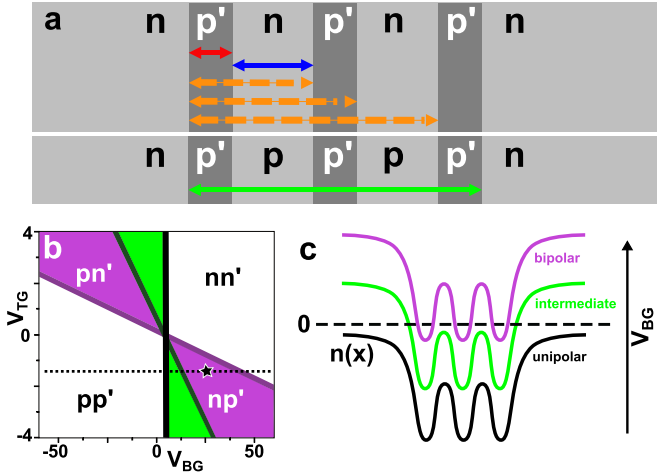}
\caption{\textbf{Resonances and resistance behavior in a graphene multibarrier system}. (a) Schematic of the Fabry-P\'erot cavities in the multibarrier system of sample A, region A3 in the bipolar and intermediate regime, respectively. Increased reflectivity for ballistic charge carriers in the bipolar regime yields resonances (colored arrows) at the pn junctions. (b) Sketch of the expected resistance hourglass-like pattern for the samples presented in this paper, where the different colors stand for the different transport regimes. The primed letters stand for the charge carriers in the barrier, the unlabeled for those in the well of the lattice. The bold black line depicts the charge neutrality line (Dirac point) of the area without top gate. The dark green line stands for the charge neutrality line of the wells and the dark purple for the one of the barriers. The green area is denoted as intermediate bipolar regime in the text. The black star points out the situation shown in (a). The schematic of according charge carrier density is shown in (c), where the three different regimes are plotted with respect to charge neutrality.}
\label{pic:multibarrier}
\end{figure}

 Using both the planar Si$^{++}$ back gate and the patterned Pd top gate electrode, the charge carrier density can be tuned both locally and globally and hence an array of potential barriers is created.
 Fig.\ \ref{pic:multibarrier}(a) shows a schematic of the Fabry-P\'erot cavities in the exemplary multibarrier system of three-striped sample region A3. For simplicity, we address the graphene cavity under the top gate stripes as ``barrier'', the one between neighboring top gates as ``well''. The region without top gate is referred to as the ``ungated'' area, that is the area between the gated region of barriers and wells and the probe contacts of the Hall bar. The regions of different charge carrier densities are depicted by different grayscale in the figure.

In the fully phase coherent bipolar transport regime of Fig.\ \ref{pic:multibarrier}(a), Klein-collimation yields an increase of reflectivity for obliquely incident charge carriers at the multiple pn or np junctions, hence, the finesse of the cavities is greatly increased. The reflections in the barrier cavities are highlighted with a red arrow, the ones in the wells are marked blue. The possible reflection traces kept in orange are less relevant due to high reflectivity of the bipolar junctions in the transmission path for oblique incidence. The green trace (i.e. the intermediate regime, see below) is more important to note due to a mismatch of potential in the gated area and the ungated area, as can be explicitly seen in the sketched corresponding charge carrier density profile in Fig.\ \ref{pic:multibarrier}(c). This difference can be attributed to the impact of the top gate stripes on the well area due to electric field broadening. The oscillations in this larger cavity will play a role later on when we come to explain the difference between measurement and simulation.

Plotting the resistance with respect to top- and back gate voltage yields four quadrants, depending on different combinations of charge carrier concentration of barriers, wells and ungated area [Fig.\ \ref{pic:multibarrier}(b)]. The  schematic color map features an hourglass-shaped pattern of high resistance, that in general can be attributed to high resistivity of the bipolar pn$^\prime$ or np$^\prime$ junctions of the gated area, where the primed charge carrier concentrations stand for the area under the top gate stripes, the unprimed ones for the well areas, respectively. This region of bipolar transport is coarsely depicted by purple and will be analyzed for the case of ballistic transmission later on in more detail. The white areas of the color map stand for the unipolar pp$^\prime$ or nn$^\prime$ low resistance transport regime, where the barrier and well charge carrier concentrations are of the same sign. The narrow green area of higher resistance is an intermediate bipolar regime, where the gated area is still unipolar, but already enclosed by the ungated area of different charge polarity.

In Fig.\ \ref{pic:multibarrier}(c) we address the different situations of the color map and plot schematics of the charge carrier density profiles $n(x)$ with respect to charge neutrality. The black curve stands for the unipolar regime, where a negative top gate voltage creates a modulation of the potential profile. Note the exaggerated mismatch of carrier density heights in the gated and ungated area. Now following the black dotted cut in the color map of Fig.\ \ref{pic:multibarrier}(b) by increasing the global back gate voltage from negative values towards positive ones, the carrier density is moved upward until it fulfills the charge neutrality condition for the ungated area. That is the bold black vertical line in the color map, separating the unipolar and bipolar regime. For pristine or homogeneously doped MLG, this line of neutral charge should intersect with the Dirac point of the locally gated area, that is the center of the hourglass pattern where the dark green and dark purple border lines intersect.

Further increase of voltage induces electrons as charge carriers in the ungated area and one enters the narrow intermediate regime [green curve in Fig.\ \ref{pic:multibarrier}(c) and green area of Fig.\ \ref{pic:multibarrier}(b), respectively], followed by another crossing (dark green diagonal line of the colorplot). That is the charge neutrality line of the wells, separating the intermediate and bipolar regime. Another increase lifts the modulated potential into the bipolar regime, where the charge carriers in the barriers and in the wells are of different sign [light purple curve and area respectively, same situation as in Fig.\ \ref{pic:multibarrier}(a)]. A raise of back gate voltage finally leads to the crossing of the barrier charge neutrality line, that is the dark purple line of the color map, where the back- and top gate induced displacement fields cancel each other and the charge carrier density is minimal under the stripes.
Concluding the trip following the dotted line, one ends up in the unipolar nn$^\prime$ situation, where the overall charge carrier density modulation is still unchanged, whereas the charges in barrier and well are of the same sign again.

\subsubsection{Transport measurement on device A}

The mobility of sample A is approximately 7000 cm$^2$V$^{-1}$s$^{-1}$, following a standard capacitor model. This implies a mean free path for charge carriers $l_m$, exceeding the effective bipolar barrier width, as well as the effective well width (e.g. $l_m\approx 100$ nm at $V_{BG}=20$ V), so that transport is ballistic over one stripe, but not over the entire lattice.

\begin{figure*}
\includegraphics[height=6cm]{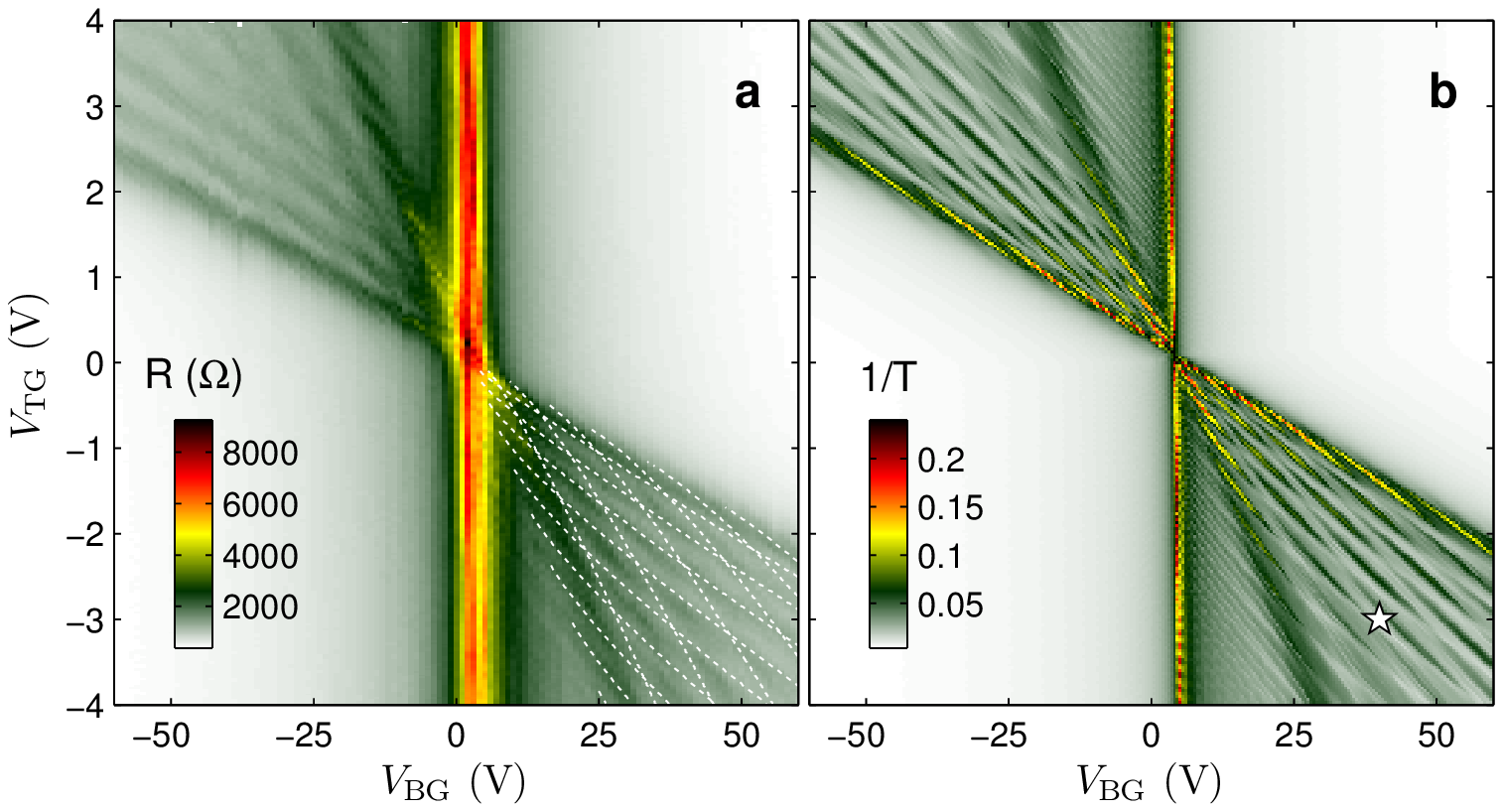}
\includegraphics[height=6cm]{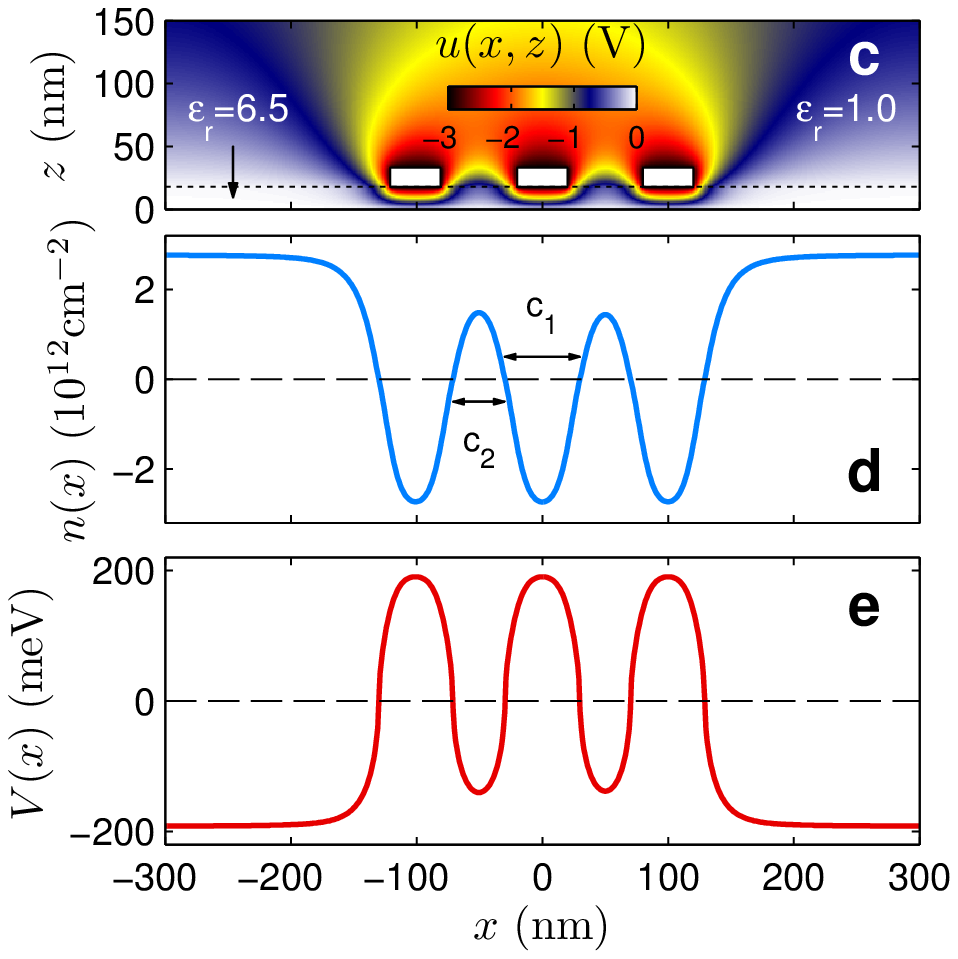}
\caption{\textbf{Resonance pattern in the bipolar regime}. Color-coded plot of (a) the four point resistance for region A3 with $w\approx 25$ nm stripe width and $a_{A3}\approx 100$ nm periodicity, and (b) the inverse of the calculated transmission function in fully phase coherent limit. A pronounced resistance pattern occurs in the bipolar regime, arising from superposition of barrier and well resonance. (c) Electrostatic simulation for the top gate contribution to the carrier density in graphene, which locates at $z=0$, at $V_{\rm TG}=-3$ V as an example. Together with the uniform contribution from the back gate set to be $V_{\rm BG}=40$ V as an example [$(V_{\rm BG},V_{\rm TG})=(40,-3)$ V point is marked by the white star in (b)], the carrier density profile and the corresponding local energy band offset profile are sketched in (d) and (e), respectively. The white dashed lines sketched in (a) are resonant voltage contours $V_{\rm TG}^{{\rm res},j}(V_{\rm BG})$ and $V_{\rm BG}^{{\rm res},j}(V_{\rm TG})$ obtained by numerically solving Eq.\ \eqref{resonance condition} for cavity $c_1$ and $c_2$ [marked in (d)], respectively; see text.}
\label{pic:G_a3_colorplot}
\end{figure*}

Figure \ref{pic:G_a3_colorplot}(a) shows one of the main results of this work, the color-coded resistance map of region A3 with respect to the two gate voltages. The four-point resistance $R_{\rm A3}$ is taken from the voltage drop across the multibarrier region A3 with a pitch of $a_{\rm A3}=100$ nm. The relatively large free space between the locally gated area and the probe contacts yields an intense, vertical, top gate-independent resistance maximum ($R\approx 10$ k$\Omega$ at $V_{\rm BG}=3$ V), that is the Dirac point of the ungated area. In the bipolar transport regime, pronounced resistance oscillations, coming from barrier and well resonance, form the aforementioned hourglass-like pattern. The white dashed lines in the lower right quadrant refer to oscillations originating in these two different cavities and will be addressed in Sec. II B 4 in more detail. The dark diagonal charge neutrality line of the well area is clearly observable and separates the oscillating bipolar region from the narrow intermediate bipolar region [cf. green area of Fig.\ \ref{pic:multibarrier}(b)]. Unlike the ungated area around the leads, where the charge carrier density is solely controlled by the back gate voltage, the top gate additionally influences the potential of the well area due to effective electric field broadening, coming from electrostatic edge effects at the Pd electrode stripes. This leads to the intermediate situation, where the polarity of barrier and well is similar, but the one of the ungated area is already different.

Note also, that the capacitive coupling of the top gate compared to the back gate for both locally gated areas A3 and A5 of device A is 26-times higher, which can be attributed to high ratio of dielectrics thickness and disparity in dielectric constant of the two oxides.

\subsubsection{Transport simulation}

To identify the source of the Fabry-P\'erot interference patterns observed in Fig.\ \ref{pic:G_a3_colorplot}(a), ballistic transport calculations were performed using the Green's function formalism in the phase coherent, clean limit. Following the SEM image of the region A3 shown in Fig.\ \ref{pic:device}, we consider a tight-binding model Hamiltonian (with hopping strength $t$),
\begin{equation}\label{H0}
H_0=-t\sum_{\langle m,n\rangle}c_m^{\dagger}c_n
\end{equation}
for a 1 $\mu$m wide and 600 nm long armchair graphene ribbon. The summation in Eq.\ \eqref{H0} runs over all the lattice site indices that are nearest neighbors to each other, $\langle m,n\rangle$. The diagonal elements of the model Hamiltonian are modulated by a realistic local energy band offset profile that is linked to the experimental parameters of gate voltages. For simplicity, we assume such a band offset profile to vary only with coordinate $x$ along the transport direction, and perform two-dimensional ($x$-$z$) electrostatic simulation to obtain the carrier density profile $n(x)$, from which, combined with the quantum capacitance model,\cite{Fang2007,Liu2013} the gate-dependent energy band offset profile $V(x)$ can then be deduced.

Instead of fully taking into account the Hall bar structure used in the 4-point measurement, we simplify the transport problem to a 2-terminal structure, which is proved to be enough to capture the main experimentally observed features. Hence together with the self-energies $\Sigma_L$ and $\Sigma_R$ due to the coupling to the left and right leads, the full model Hamiltonian reads
\begin{equation}\label{H}
H=H_0+\sum_{n}V(x_n)c_n^{\dagger}c_n+\Sigma_L+\Sigma_R,
\end{equation}
where the summation runs over all the lattice site indices and $x_n$ is the $x$-coordinate of the $n$th lattice site.

With the model Hamiltonian \eqref{H}, the transmission function $T$ is computed by the recursive Green's function method.\citep{Wimmer2009} Figure \ref{pic:G_a3_colorplot}(b) shows the inverse of the calculated transmission function, which agrees qualitatively well with the measured resistance map shown in Fig.\ \ref{pic:G_a3_colorplot}(a). Such a good agreement relies closely on the optimal electrostatic model that generates the correct gate voltage dependence of the local energy band offset applied to the diagonal elements of the model Hamiltonian. The geometry of the adopted electrostatic model is sketched in Fig.\ \ref{pic:G_a3_colorplot}(c).

To obtain a close match between the experimental results and the numerical simulations, the oxide thickness and Pd stripe width (fabrication design values 15 nm and 25 nm, respectively) were slightly modified to 18 nm and 40 nm, respectively. Note also that the back gate contribution is assumed to be uniform, so that the standard capacitor model can be used. Furthermore a uniform $p$-type chemical doping concentration of $n_0=-2\times10^{11}\operatorname{cm}^{-2}$ is considered, which is deduced from the slightly shifted Dirac point in the ungated region [vertical thick line in Fig.\ \ref{pic:G_a3_colorplot}(a)].

For details about simulating the gate-induced carrier density in graphene, see, for example, Ref.\ \onlinecite{Liu2013a} for a tutorial introduction using the \texttt{pdetool}\cite{pde} of {\sc Matlab}. Alternatively, one can as well choose the free automated finite element simulator--the FEniCS project,\citep{FEniCS} which is adopted, combined with the free mesh generator--GMSH,\cite{gmsh} in the present work.

\subsubsection{Analysis on the Fabry-P\'erot interference patterns}

Figures \ref{pic:G_a3_colorplot}d and e illustrate as an example the carrier density profile $n(x)$ and the corresponding energy band offset profile $V(x)$, respectively, at gate voltages $(V_{\rm BG},V_{\rm TG})=(40,-3)$ V, as marked by the white star in Fig.\ \ref{pic:G_a3_colorplot}b. We next attribute the different set of Fabry-P\'erot interference patterns to the resonance in specific cavity regions. For this, we first label the cavity under the top gate (barrier) $c_1$, and the region between two neighboring top gates (well) $c_2$. Taking the $n(x)$ sketched in Fig.\ \ref{pic:G_a3_colorplot}d for example, we may regard $c_1$ as the region $-30\operatorname{nm}\lesssim x\lesssim 30\operatorname{nm}$ and $c_2$ as $-70\operatorname{nm}\lesssim x\lesssim -30\operatorname{nm}$. Within cavity $c_i$, the resonance condition can be qualitatively written as
\begin{equation}\label{resonance condition}
2\int\limits_{c_i}\sqrt{\pi|n(x)|}dx=2j\pi,\qquad j=1,2,\cdots,
\end{equation}
where $\sqrt{\pi|n(x)|}=k_F(x)$ is the position dependent wave number. The left hand side of Eq.\ \eqref{resonance condition} can be understood as the phase difference between directly transmitted and twice reflected electron waves. When such a phase difference is an integer multiple of $2\pi$,  constructive interference is expected.

Since $n(x)$, and hence the cavity size, vary with the gate voltages, Eq.\ \eqref{resonance condition} has to be solved numerically. At a fixed back gate voltage, there exists a discrete number of top gate voltages that satisfy the resonance condition \eqref{resonance condition} when solving for the cavity $c_1$. Sweeping the back gate voltage in the bipolar range, one obtains the resonant top gate voltage contours as a function of back gate voltage, $V_{\rm TG}^{{\rm res},j}(V_{\rm BG})$, which form a group of lines that are roughly parallel to the diagonal charge neutrality line of the barriers, that is the dark purple boundary discussed in Fig.\ \ref{pic:multibarrier}(b). Likewise, one can solve Eq.\ \eqref{resonance condition} for the cavity $c_2$ to obtain $V_{\rm BG}^{{\rm res},j}(V_{\rm TG})$ contours, which form a group of lines roughly parallel to the steep, dark diagonal charge neutrality line of the wells, shown in Fig.\ \ref{pic:G_a3_colorplot}(a) or (b), also schematically depicted by the dark green border in Fig.\ \ref{pic:multibarrier}(b).

Sketching the $V_{\rm TG}^{{\rm res},j}(V_{\rm BG})$ and $V_{\rm BG}^{{\rm res},j}(V_{\rm TG})$ contours on the resistance map of Fig.\ \ref{pic:G_a3_colorplot}(a) (inset of dashed white lines), we find a very good agreement of the periodicity of the interference patterns. This enables us to correctly attribute the different origins of the different sets of interference patterns to the resonance in different cavities $c_1$ (barrier) and $c_2$ (well).

\subsubsection{Discussion}

Despite the good agreement between the measurement [Fig.\ \ref{pic:G_a3_colorplot}(a)] and simulation [Fig.\ \ref{pic:G_a3_colorplot}(b)], a closer look at the latter indicates that the interference patterns in the ideal calculation are more complicated and more pronounced than the former in the bipolar regime due to phase coherence assumed over the whole area. Addressing the cavity picture of Fig.\ \ref{pic:multibarrier}(a), the fringes result primarily from superposition of barrier- (red) and well resonance (blue). Moreover, the simulated map of Fig.\ \ref{pic:G_a3_colorplot}(b) shows fringes with a shorter oscillation period, arising in the intermediate region. These can be traced back to resonances in a larger cavity [cf. green trace in Fig.\ \ref{pic:multibarrier}(a)]. In the experiment, we are not able to see the short period oscillations in this particular region due to the fact that the length of the intermediate cavity exceeds the mean free path of the sample. Instead, a blurred area of higher resistance is observed and can be attributed to diffusive transmission in the intermediate bipolar regime.

In order to get an insight into the role of a finite mean free path and phase coherence length, we successively reduced the number of barriers in the simulation, while retaining the mismatch of inner and outer well potential (not shown). Comparing the measured and simulated resistance plot qualitatively, we estimate the ballistic length to 1.5 times the effective barrier width, as only the oscillations coming from barrier resonance are clearly pronounced [Fig. \ref{pic:G_a3_colorplot}(a)]. Evidently, ballistic transmission in the range of the barrier width is essential to evoke resistance oscillations, on the other hand, phase coherence over the whole top gated area is not required to reproduce the general behavior. Hence, the multibarrier system can be understood as independent FP cavities stringed together, giving rise to a single barrier pattern in the locally ballistic limit.

\subsection{Devices B and R}

\begin{figure*}[t]
\includegraphics[width=0.95\textwidth]{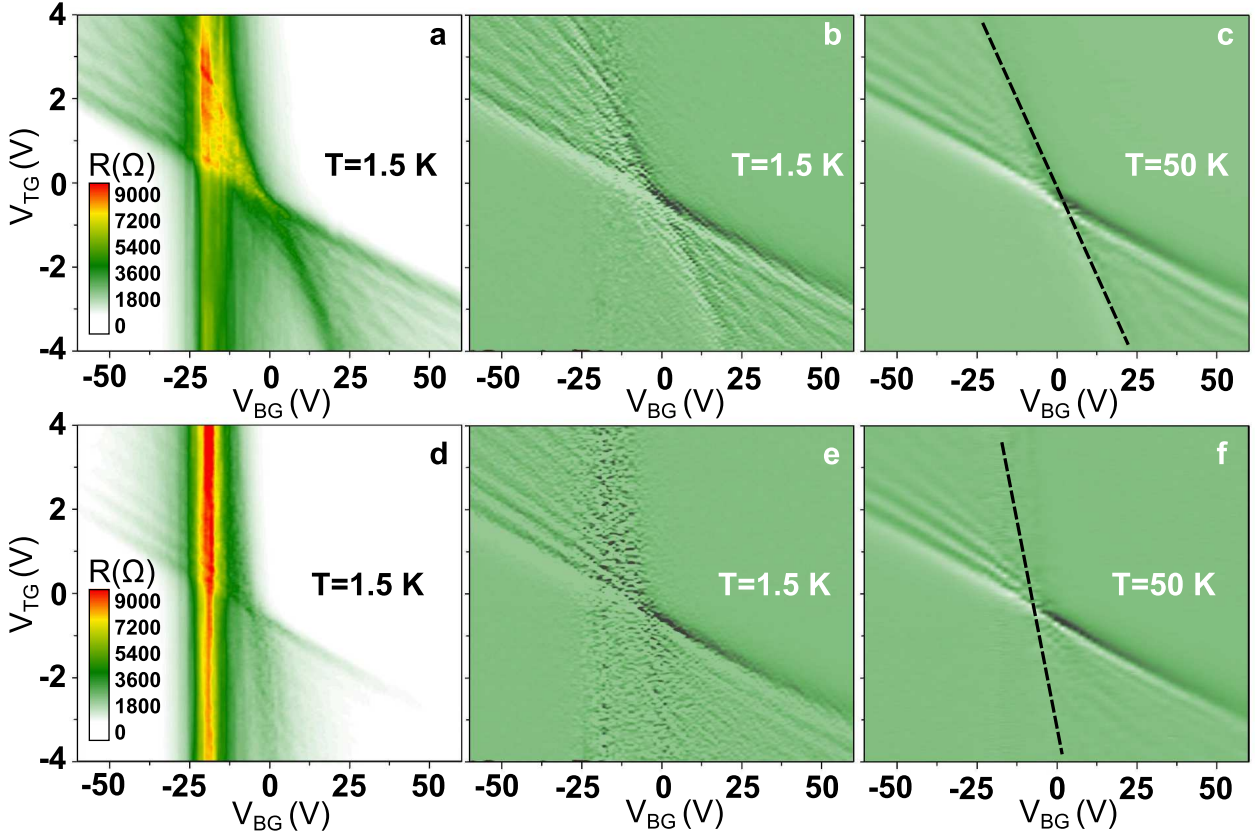}
\caption{\textbf{Resistance ocillations of sample B.} (a) Color map of resistance in region B5 (five stripes) and its derivative with respect to the top gate voltage  (b). A pattern with barrier and well resonance contribution is observable, the vertical charge neutrality line of the ungated area is filtered. (c) At $T=50$ K, the superposition fades out and one obtains a softened single barrier resonance pattern. Similar data for region B2 is plotted in (d)-(f).}
\label{pic:b5_derivatives}
\end{figure*}

Next, we discuss device B, that shows similar behavior to device A with respect to mobility ($\mu_B \sim 6300 - 8300$ cm$^2$V$^{-1}$s$^{-1}$) and resonance pattern [Fig. \ref{pic:b5_derivatives}]. The top gate voltage-independent charge neutrality line of the ungated area though does not coincide with the ones of the hourglass-like pattern of B5 [Fig.\ \ref{pic:b5_derivatives}(a)]. The ungated region appears to be n-doped (DP at $V_{\rm BG} = -20$ V, whereas the top gated zone B5 shows a lower dopant level (DP for $V_{\rm TG} = 0$ V  at $V_{\rm BG} = -4$ V). Interestingly, this inhomogeneity is consistent for both top gate areas, B2 and B5, that are spatially separated and thus the inhomogeneity can only be attributed to some preparation step of the top gate oxide and electrodes that changed the intrinsic doping at the different sites. Keep in mind that the spacing between the top gate stripes of the regions B2 and B5 is different, $a_{\rm B2}=200$ nm and $a_{\rm B5}=100$ nm, whereas the width of the top gated area (barrier) is similar $\approx 25$ nm [cf. SEM-picture in Fig. \ref{pic:device}(a)].

The impact of the top gates on the charge carrier density in the region between two stripes is lower for region B2 and therefore the potential mismatch of the well and ungated region is smaller. This results in different slopes of the well charge neutrality lines in the resistance maps [dashed slopes in Fig.\ \ref{pic:b5_derivatives}(c) and (f)]. One can assume, that for the larger well cavity of B2 (where $a_{\rm B2}>l_m$), ballistic interference is suppressed. On the other hand, one can expect a superposed resonance pattern of barrier and well cavities for B5, as  $a_{B5}\lesssim l_m$, similar to sample A3. Fig.\ \ref{pic:b5_derivatives}(b) shows the numerical derivative of the resistance pattern of B5 with respect to the top gate voltage. In this plot, superposition of resonances can be resolved, but the bright charge neutrality line of the ungated region is filtered. The derivative of the resistance in region B2 [Fig.\ \ref{pic:b5_derivatives}(e)] only features the pronounced parallel resonance peaks coming from single barrier resonance.

\FloatBarrier

\begin{figure}[h]
\includegraphics[width=\columnwidth]{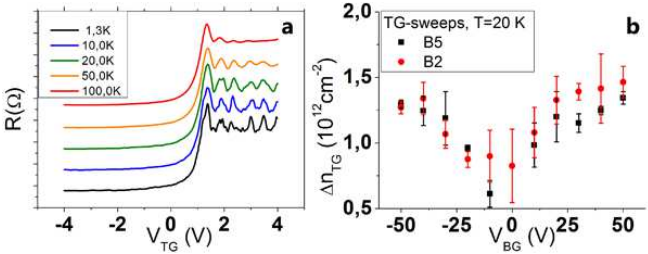}
\caption{\textbf{Temperature dependence and resistance oscillation period of sample B}, (a) Temperature development of a top gate sweep. Sweep performed at a back gate voltage $V_{\rm BG} = -40$ V in region B5. The curves are offset by 500 $\Omega$ for clarity. (b) Comparison of the oscillation periods $\Delta n_{\rm TG}$ for sample regions B5 and B2 for top gate voltage sweeps at different back gate voltages at $20$ K. $\Delta n_{\rm TG}$ is the average spacing between two resistance peaks of a single top gate sweep as depicted in (a), with respect to change in charge carrier density.}
\label{pic:B_data}
\end{figure}

We performed temperature-dependent measurements on device B and find that the FP oscillation amplitude is decreasing with higher temperatures $T$ [cf. Fig.\ \ref{pic:B_data}(a)]. As expected, the period of resistance oscillation with respect to charge carrier density remains unchanged. The features originating from superposition of barrier and well resonance at $1.5$ K are successively suppressed  at higher $T$ that is due to temperature activated dephasing and loss of ballistic transmission. As the phase coherence length decreases with $T$, only the single barrier resonance survives, resulting in a softening of the resistance pattern.

Comparison of different top gate voltage sweeps at 20 K for regions B5 and B2 [Fig.\ \ref{pic:B_data}(b)] yields that the oscillation period of the remaining pronounced resonance fringes with respect to the charge carrier density is pretty similar for both regions. This, together with the fact that the resonance pattern remains observable at higher temperatures, clearly proves local ballistic Fabry-P\'erot interference originating in the barriers.

\begin{figure}[h]
\includegraphics[width=\columnwidth]{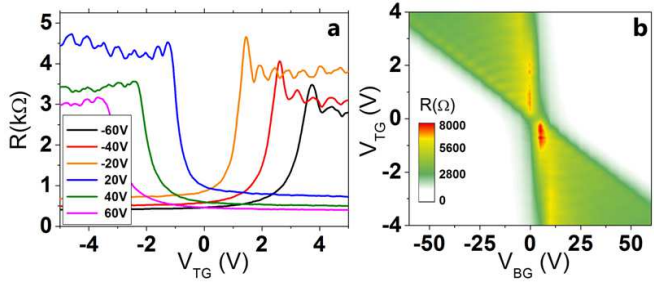}
\caption{\textbf{Resistance oscillations of sample R8.} R8 has been fully covered by top gate stripes ($a_{\rm R8}=200$ nm) and shows for a sample area of 8 stripes similar behavior to devices A and B. The resistance of single top gate sweeps at different back gate voltage and at $T = 1.5$ K (a) feature clear oscillations. The resistance map shows the general hourglass-like pattern, separating the map into four quadrants (b).}
\label{pic:R03_data}
\end{figure}

Finally we briefly discuss our last sample R. Again, the electronic properties are very similar to the other two devices A and B, however it has been fully covered by a lateral top gate electrode array of $200$ nm lattice constant. In Fig.\ \ref{pic:R03_data}(a), we observe pronounced FP oscillations at $T = 1.5$ K for a sample region of 8 stripes (R8). While the large steps in $V_{\rm BG}$ do not allow resolving the fine structure of this pattern, we once again recognize the characteristic hourglass shape in Fig.\ \ref{pic:R03_data}(b), but do not encounter the intermediate bipolar regime [green area in Fig.\ \ref{pic:multibarrier}b], as it is exclusive to devices A and B with local gating. Here, the top gate electrode lattice covers the whole transport channel. Hence, there is no area without top gate stripes and the potential shows no mismatch in this case.

\FloatBarrier

\subsection{Towards superlattices}

\begin{figure}[h]
\includegraphics[width=\columnwidth]{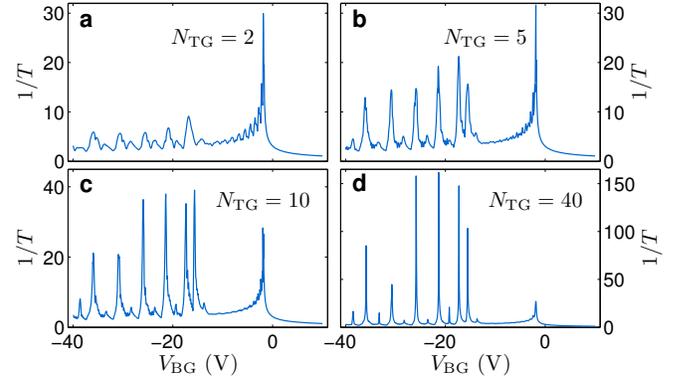}
\caption{\textbf{Impact of superlattice on a fully phase coherent graphene system.} The figures (a)-(d) depict the calculated inversed transmission for a single back gate sweep for different numbers of barriers $N_{\rm TG}$. The peak of the original Dirac point occurs at $V_{BG} \approxeq -2$ V and, with increasing $N_{\rm TG}$, gets surpassed by resistance peaks originating in destructive interference.}
\label{pic:newDP}
\end{figure}

The analysis of our data lets us finally conclude that there is no evidence of an artificial band structure and thus a superlattice effect in our multibarrier system. This stands in contrast to the argumentation of Ref. \onlinecite{dubey2013tunable}, where a similar device with an even higher number of top gate stripes was investigated. Our calculations for a fully phase coherent system of $N_{\rm TG}$ top gate stripes yield growing of resistance peaks due to the formation of new touching points of valence and conductance band as the band structure gets folded, giving rise to the opening of small energy band gaps. In the FP picture, in the case of only a few periods, the destructive interference gives rise to these resistance peaks and these peaks do not exceed the one at the usual Dirac point. This is consistent with all data extracted from our samples. In Figs.\ \ref{pic:newDP}(a)-(d) we plot the calculated inversed phase coherent transmission of a back gate sweep at $V_{\rm TG}=3$ V for various numbers of top gate stripes, showing a global Dirac point at $V_{\rm BG}\approxeq -2$ V. When $N_{\rm TG}$ exceeds $\sim 10$, the heights of the resonance peaks start to be of the same order or even higher than that of the usual Dirac point. One should see the resistance at those anti-crossings as diverging in comparison to that at the usual Dirac point, however, the miniband structure should be extremely sensitive to inhomogeneity. This is why the mean free path and phase coherence length appear to be a limiting factor for the emergence of mini band gaps and diverging resistance peaks. Hence a combination of high-mobility graphene on a substrate (i.e. hBN/graphene heterostructures) and small top gate lattice periods seems to be the most obvious option in order to pave the way to sizable superlattice effects in graphene.

\FloatBarrier

\section{Conclusion}

In summary, we fabricated top gated graphene multibarrier systems with different number of bipolar potential barriers and spacing. We observe highly pronounced resistance oscillations in the bipolar transport regime in each of our devices, which can be attributed to Fabry-P\'erot resonance in the ballistic limit of one barrier or well length. In the locally gated samples, we encounter three charge neutrality lines that come from different potentials in the gated and ungated areas. Moreover, we consistently reproduced our measurements in a transport simulation, employing a realistic potential profile, extracted from electrostatics and a quantum capacitance model.

\section*{Acknowledgments}
Financial support by the Deutsche Forschungsgemeinschaft (DFG) within the programs GRK 1570 and SFB 689 and by the Hans B\"ockler Foundation (F.T.) is gratefully acknowledged.

\end{document}